\documentclass[]{tPHM2e}

\begin{document}
\doi{10.1080/14786435.20xx.xxxxxx}
\issn{1478-6443}
\issnp{1478-6435}
\jvol{00} \jnum{00} \jyear{2010} 

\markboth{Taylor \& Francis and I.T. Consultant}{Philosophical Magazine}

\title{Compaction of cereal grain}

\author
{J. Wychowaniec$^{\rm a}$ $^{\ast}$\thanks{$^\ast$Corresponding author. Email: JacekWych@hotmail.com\vspace{6pt}} , 
I. Griffiths$^{\rm b},$\\\vspace{6pt}  
A. Gay$^{\rm b}$,  
A. Mughal$^{\rm a}$\\\vspace{6pt} $^{\rm a}
${\em{Institute of Mathematics and Physics, Aberystwyth University, Penglais, Aberystwyth, Ceredigion, Wales, SY23 3BZ, United Kingdom}}\\\vspace{6pt}
 $^{\rm b}${\em{Institute of Biological, Environmental and Rural Sciences, Aberystwyth University, Gogerddan, Aberystwyth, Ceredigion, Wales, SY23 3EE, United Kingdom}}\\\vspace{6pt}\received{v4.5 released May 2010} }

\maketitle

\begin{abstract}
We report on simple shaking experiments to measure the compaction of a column of Firth oat grain. Such grains are elongated anisotropic particles with a bimodal polydispersity. In these experiments, the particle configurations start from an initially disordered, low-packing-fraction state and under vertical shaking evolve to a dense state with evidence of nematic-like structure at the surface of the confining tube. This is accompanied by an increase in the packing fraction of the grain. 

\bigskip

\end{abstract}

\section{Introduction}

Cereal products are one of the most important staple foods for both humans and domesticated livestock. As a result the commerce in cereals and other food grains such as wheat, maize and rice is one of the oldest forms of trade in the world and is today an industry of vital global importance. 

One of the most ancient and commonly used specifications in wheat grading is the hectolitre weight (HLW kg hl$^{-1}$, also known as specific weight, test weight or bushel weight), which is a measure of the mass of grain that packs into a specified volume  \cite{doehlert, doehlert2, doehlert3, white}. Typically HLW is measured using a {\it chondrometer}. This is device that consists of an upper and lower cylinder, which are isolated from each other by a slide gate (a metal blade). The upper cylinder is filled with a column of grain while the lower cylinder is empty. Opening the gate allows grain to fall under gravity into the lower cylinder in a controlled fashion. Reinserting the blade levels off the excess, after which the weight of the lower cylinder is recorded. From this the HLW value of the grain can be computed. 

HLW is widely believed to be a measure of the bulk density of the grain. It is commonly used as indicator of grain quality and millability (to ensure effective milling the hectolitre weight should be above 76 kg hl$^{-1}$ for wheat and 50 kg hl$^{-1}$ for oats). As a consequence grain with a high HLW commands a higher price per ton \cite{manley}. 

The persistence of HLW as a measure of grain quality can be attributed to the fact that the chondrometer test is cheap and simple to perform. However, the fairness of the result is a hotly contested issue \cite{manley}. This is in part due to the availability of a variety of chondrometer models (of differing dimensions) including some that employ a funnel to fill the measuring cylinder. In some cases the measuring device is given a few taps to encourage the grain to settle, while in other cases it is not. As a result there can be significant variation in the HLW value depending on the device and the protocol used.

Clearly, the measured value of the hectolitre weight is also due to a combination of grain characteristics including friction, grain shape and polydispersity. As such, considerable effort has been made to characterise the shape of individual grains using optical methods (as described below). Although modern varieties of wheat have less variation in size; nevertheless, tritiacae variability still occurs \cite{gegas}. 

In contrast oat is a more complex grain and is often described by a bimodal size distribution \cite{doehlert3}. This arises from the arrangement of grains within the oat spikelet, which contains between one and three grains. Of these 80\% contain two (primary and secondary) grains \cite{doehlert3}. The primary grain is longer than the secondary grain but their widths are similar. For all cereals grains the polydispersity also depends on the  cultivar used and the environmental conditions during grain filling.

In this study we show that packing efficiency can vary widely as a function of grain handling. In particular, we investigate the compaction of a column of Firth oat grains (the most popular spring oat in the UK with a large market share) by means of mechanical vibration. We begin by simply pouring the grain into a perspex tubular container (of known dimensions), this mechanically stable configuration represents an initial loosely packed metastable state. Since thermal energies, $k_BT$ are insignificant compared to the energy required to rearrange a single grain each such metastable configuration will persist indefinitely until some external vibration is applied which moves the system into another state (i.e. no thermal averaging takes place to equilibrate the system). By shaking the column we demonstrate that a range of packing densities are realisable, where the degree of compaction depends on the shaking protocol.

On the one hand these results serve to elucidate the role of mechanical agitation in determining the bulk density, and as a result the HLW, of grain products. On the other hand they are useful in understanding grain compaction in seafaring silos. Whereby, grain level in such silos is known to drop significantly during a prolonged voyage due to the mechanical vibration generated by the ships engines and the swaying motion of the sea. 

Compaction of cereal grain is also of fundamental interest. Most granular compaction studies using mechanical agitation have focused on highly spherical glass particles \cite{knight, nowak, ben, rib, knight2, phil}. However, real world grains are much more complex and are often not monodisperse and frequently non-spherical. Firth oat grains are a perfect example of this complexity, the kernel is anisotropic and elongated in shape with the length following a bimodal distribution. 

We contrast these findings with compaction experiments using readily available rape seed, which unlike Firth grains are spherical in shape and have radii that can be described using a mono modal distribution.

\section{Materials and methods}

Here we discuss the details of our experimental procedure. We describe the method for characterising cereal grain samples, the experimental shaking apparatus and the protocol used.

\subsection{Grain Measurement}

\subsubsection{Firth}

\begin{figure}
\begin{center}
\subfigure[]{
\resizebox*{5cm}{!}{\includegraphics{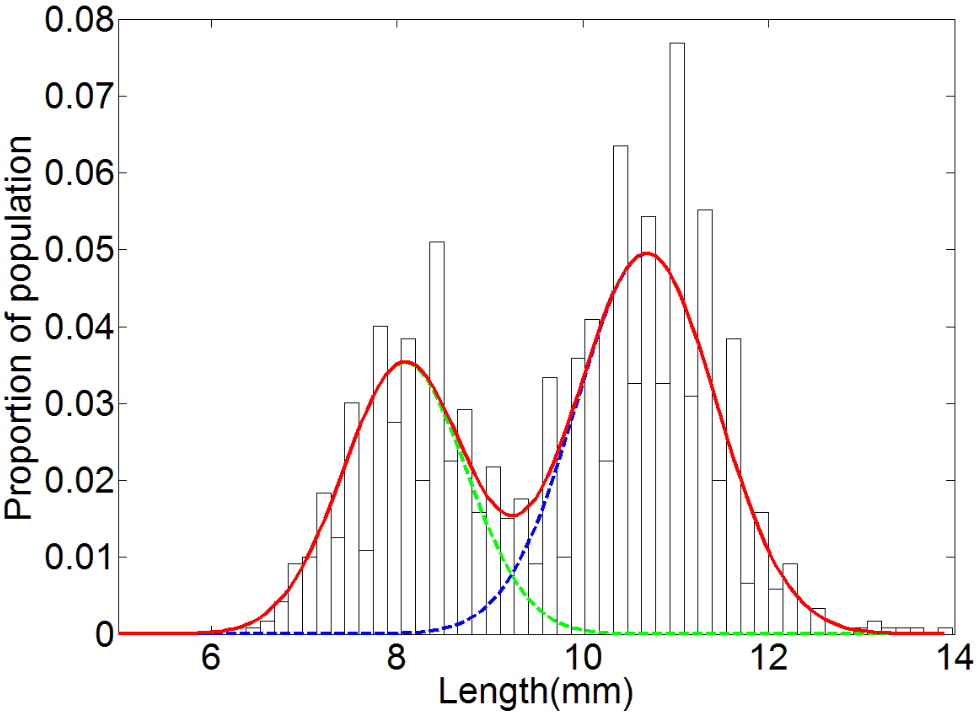}}}%
\subfigure[]{
\resizebox*{5cm}{!}{\includegraphics{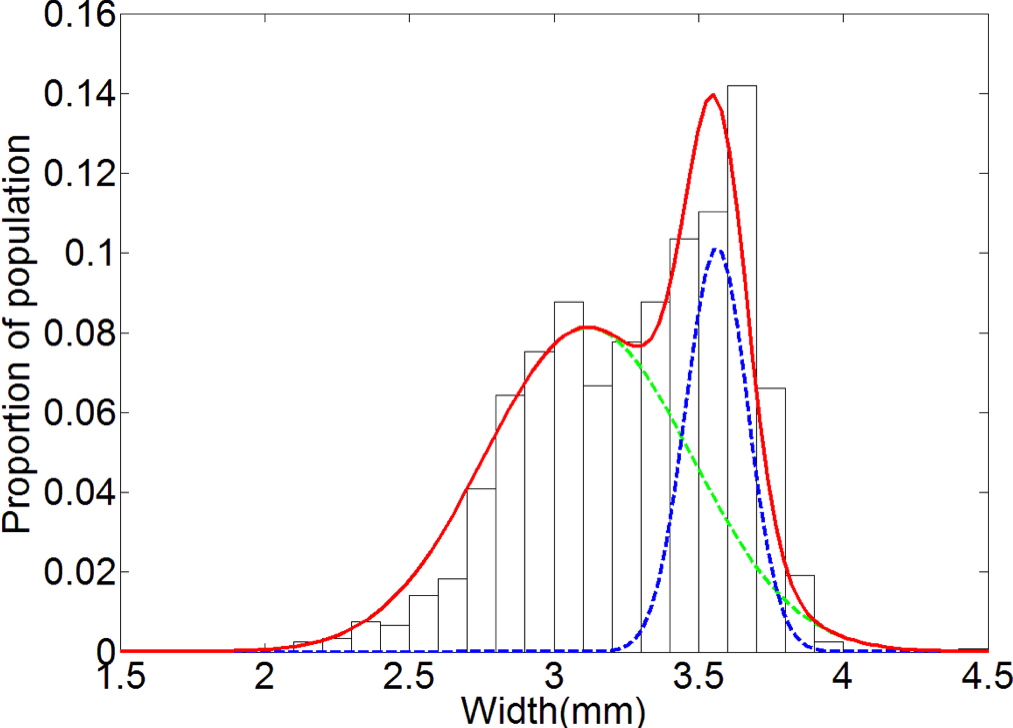}}}%
\subfigure[]{
\resizebox*{5cm}{!}{\includegraphics{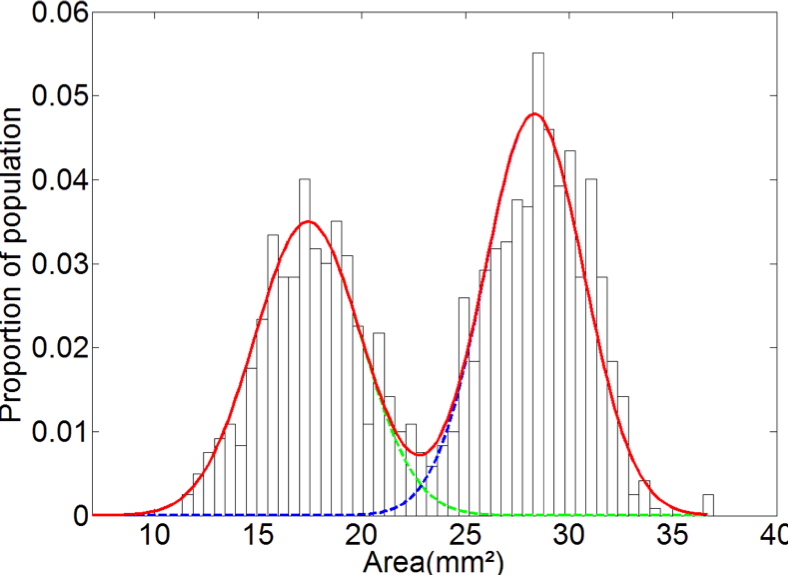}}}%
\caption{Histograms of grain length, width and cross sectional area - respectively. The green and blue lines are numerical fits to the two populations while the red line is a bimodal distribution generated by an iterative process, as described in the text.}%
\label{grain_dist}
\end{center}
\end{figure}

For these experiments the spring oat cultivar Firth was used because it is widely grown in the UK and is one of the control cultivars on the HGCA recommended list. Grain samples were taken from the 2010 spring oat trials at grown at IBERS, Aberystwyth. The size (length, width and area) of over 1000 grains was determined using a digital seed analyser (MARVIN-fine, GTA Sensorik GmbH, Neubrandenburg, Germany). The overall size distribution of the grains (d) was then represented as a bimodal distribution:
\begin{equation}
d=vf_{npd}(\mu_1, \sigma_1) + (1-v)f_{npd}(\mu_2, \sigma_2)
\end{equation}
where $\mu$ is the mean and $\sigma$ the standard deviation of the normal probability density function 
$(f_{npd})$ for the component distributions (subscripts 1 and 2) and v is the proportion in population 1. This distribution was fitted iteratively with initial values for for $\mu_1$ and $\mu_2$ set to 25\% ($e\mu_1$) and 75\% ($e\mu_2$) quartiles of the overall distribution of grain size (x). Initial values for $\sigma_1$ and $\sigma_2$ were both set to $\sqrt{(var(x)-0.25(e\mu_1-e\mu_2))^2}$ where $var(x)$ is the variance of $x$, and $v$ was always set to $0.5$. 

The results of the iterative process implemented in Matlab\textsuperscript{\textregistered}, were compared visually with the original distribution for grain width, length and area Figure~\ref{grain_dist}. Both length and area show a clear bimodal distribution, although with different proportions in the two distributions. For width there was less of a distinction, probably due to the asymmetry of the cross section of grains. The parameter estimates (Table 1) show a lower proportion of grain in the first population (the smaller, secondary) grains for both length and area. The program used and sample data are available on request from the authors.

By approximating each kernel as an ellipsoid the above data can be used to compute the average volume $V_g$ of the grains in a sample.

\begin{table}[ht]
\centering
\begin{tabular}{c c c c c c}
\hline\hline
Measurement & \shortstack{Proportion in first \\population} & \shortstack{Mean \\ of population 1} & \shortstack{Mean \\ of population 2} &   \shortstack{Standard \\ deviation of \\population 1} & \shortstack{Standard \\ deviation of \\population 2} \\ [0.5ex] 
\hline
Length $mm$&0.38&8.08&10.69&0.66&0.75 \\
\multicolumn{1}{r}{CI}&0.35-0.42&8.00-8.17&10.61-10.76&0.60-0.72&0.70-0.81 \\
Width $mm$&0.73&3.12&3.56&0.36&0.11 \\
\multicolumn{1}{r}{CI}&0.67-0.78&3.08-3.15&3.54-3.58&0.34-0.37&0.09-0.13 \\
Area $mm^2$  & 0.44 & 17.40 & 28.32 &2.57 & 2.39\\
\multicolumn{1}{r}{CI}&0.41-0.47&17.15-17.65&28.12-28.52&2.37-2.76&2.23-2.55
 \\ [1ex]
\hline
\end{tabular}
\caption{Parameters for the fitted distributions in Figure~\ref{grain_dist}, CI values are the 95\% confidence intervals of the fitted parameter immediately above}
\label{table:nonlin}
\end{table}

\subsubsection{Oil seed rape}

Measurements were also made on seeds of oil seed rape (Compass) to provide a comparative biological sample that is approximately spherical in shape. When measured on Marvin (using 1183 grains), as above, these seeds were shown to posses a unimodal distribution with mean radius 1.36 mm and standard deviation 0.076 mm (and a standard error of 0.0022 mm). Note, because it was impossible to separate width and depth of the flattened oat seeds from the two-dimensional image analysis data, and because of the limited resolution (Å0.09 mm) when measuring the oil seed rape, ancillary measurements of dimensions of over 100 seeds of each sample were made using a digital calliper resolution 0.01 mm. These calliper measurements for the oil seeds gave a mean radius of 1.04mm with a standard deviation of 0.10 mm and a standard error of 0.0082 mm.  

\subsection{Grain shaking experiment and protocol}

Our experimental procedure is as follows. An initially loose packed state is formed by pouring a fixed mass of grain into a square sided perspex tube of dimensions $3.3\textrm{cm}\times 3.3\textrm{cm} \times 35\textrm{cm}$, which is attached to a wooded base. Using a ruler the height of the grain is measured on all four sides. We take the average of these four readings to be the height of the grain column $h_c$. 

The mass of 1000 grains of a given cereal is a well known agricultural measure known as the thousand grain weight (TGW) and has a value of 38.42g ($\pm$ 0.33g) for Firth. Thus knowing the mass of a grain sample it is possible to compute a rudimentary estimate of the number of grains in the sample $N$. From this we can compute the average volume fraction (packing density) of the grain,
\[
V=\frac{NV_g}{V_c},
\]
where $V_c=V_c(h_c)$ is the volume of the tube that is filled with grain and depends only of the height of the grain.

\begin{figure}
\begin{center}
\resizebox*{5cm}{!}{\includegraphics{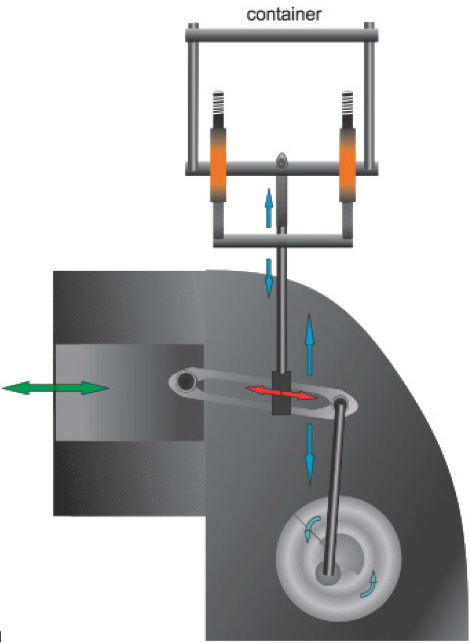}}
\end{center}
\caption{A schematic diagram of the experimental setup. The amplitude of the shaking motion can be varied by adjusting the position of a movable block (green arrows).}
\label{esetup}
\end{figure}

The base, and the attached column, can be made to oscillate rapidly in the vertical direction by the action of a DC stepper motor. The motor is connected to a cam which drives a shaft attached to an adjustable lever arm, as shown in Figure~\ref{esetup}. By adjusting the length of the lever arm it is possible to change the amplitude of the vertical oscillations. For the purposes of our experiments we fix the amplitude of oscillations to be $6\textrm{mm}$ and instead vary the frequency. An unloaded stepper motor is expected to have a one-to-one relationship between rotation speed and driving frequency. However in our case we measure the frequency of the motor using an optical sensor as an extra check; this is to account for the effect of the reaction force on the motor due to a violently shaking column of loose grains. 

We begin by shaking the sample with the maximum frequency of oscillations attainable with our motor, this corresponds to the most vigorous shaking achievable. After 300 seconds we stop the experiment and measure the height of the grain column and compute the corresponding volume fraction. After this, the oscillation frequency is reduced slightly and the whole procedure is repeated again for 300 seconds (i.e. the frequency is reduced after each 300s period of shaking). By gradually reducing the frequency we anneal the sample from a highly energetic state (in which we observe diffusion of grains across the length of the column) to a low energy state dominated by local rearrangements. Our expectation is that such a sweep will reveal the most effective driving frequencies for compacting the sample.
 
\subsection{Experimental results}

\begin{figure}
\begin{center}
\resizebox*{10cm}{!}{\includegraphics{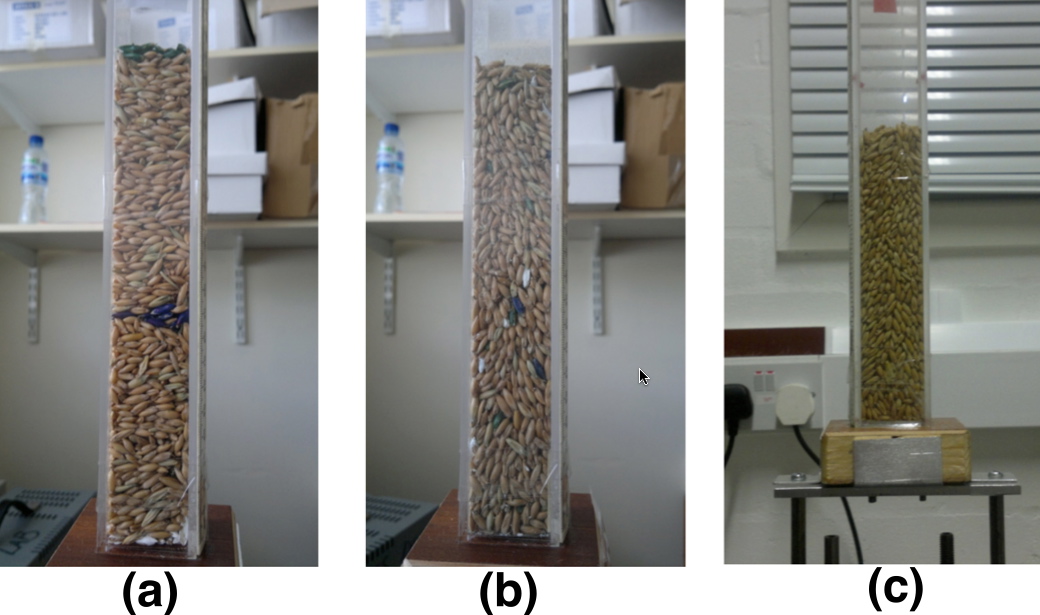}}
\end{center}
\caption{Snap shots of a column containing Firth oat grains. {\bf (a)} the initial loose packed state formed by pouring grain into the column. Oat grains at the bottom are coloured white, blue grains in the middle and green at the top. {\bf (b)} The system after shaking for 300 seconds at maximum frequency. {\bf (c )} the system at the end of the experiment, with a volume fraction of approximately 0.8.}
\label{snaps}
\end{figure}

A series of experiments have been performed on 2 types of grains - oilseed rape grain and Firth grain using the above protocol. In both cases, a general increase in packing density is observed, however this effect is more pronounced in Firth grains (elongated and bimodal) than oilseed grains (spherical and mono-modal).

Six replicated experiments were performed using about 120g of Firth grain. An example of the initial loose packed state formed by pouring grain into the column is shown in Figure~\ref{snaps}a, where we have painted grains at the bottom white and similarly coloured grains at the middle (blue) and top (green). Note, that in this initial state when the grains are poured into the sample they come to rest so that, on average, their semi-major axes are perpendicular to the axis of the confining tube.

\begin{figure}
\begin{center}
\resizebox*{13.5cm}{!}{\includegraphics[]{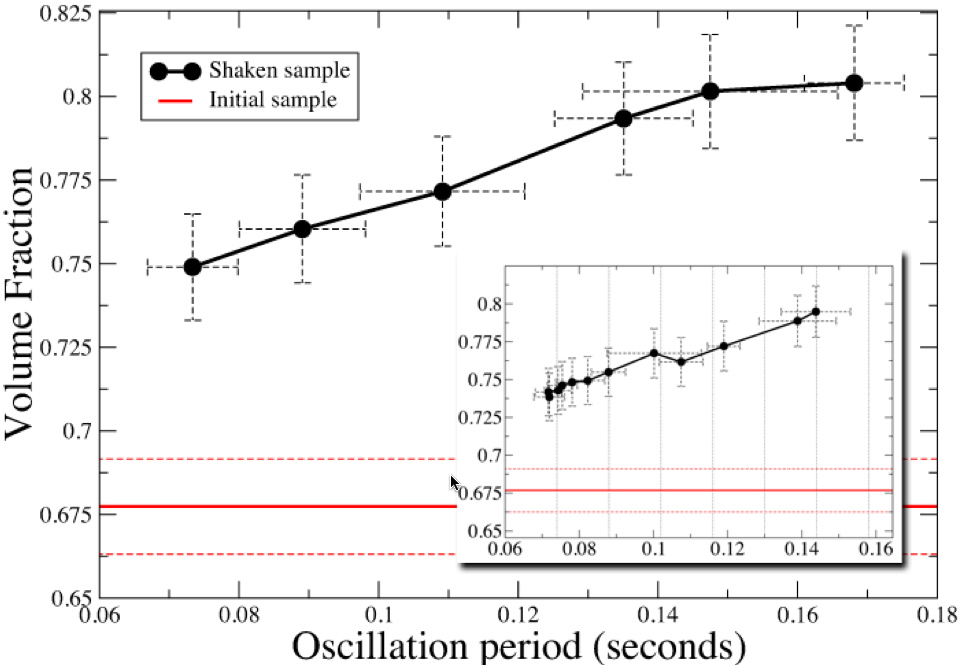}}
\end{center}
\caption{Averaged volume fraction of Firth grains as a function of shaking. The plot shows the change in volume fraction after shaking the sample for 300 seconds using oscillations of the indicated period (or inverse frequency). Six experiments were performed for a column containing approximately 120g of Firth oats. The red line indicates the volume fraction of the initial loose packed sample (red dash lines are the experimental error in this measurement). The inset shows one of the six experiments on its own. The data from each of the six runs is binned into the regions indicated on the inset. By averaging the data, from all of the six runs, into these bins an averaged volume fraction is obtained (main graph). In all cases there is a dramatic compaction after the first run (performed using the maximum frequency of the motor) subsequent densification seems to be linear and saturates at low shaking frequencies. Error bars for each measurement are also shown.}
\label{VF}
\end{figure}

We perform the experiment using the protocol outlined above, the results are shown in Figure~\ref{VF}. The inset shows the volume fraction for one of the six experiments, as a function of the average oscillation period (i.e. inverse frequency). The heavy red line indicates the volume fraction of the initial sample (i.e. before shaking), while the red dashed lines indicate experimental error in this measurement. Each data point indicates the volume fraction as obtained after each 300 second period of shaking, as a function of the average oscillation period (i.e. inverse frequency). 

Thus, for one of the experiments (shown in the inset) the initial volume fraction of the loose packed packed sample is 0.677 but after 300s of shaking, at a frequency corresponding to 0.072 seconds per oscillation, the grain column compacts to a volume fraction of 0.738. The sample is then shaken for 300s at a frequency corresponding to an oscillation period of 0.0717 seconds per oscillation and is found to compact to a volume fraction of 0.742 - and so on until the oscillations are so slight as to have a negligible effect on the grain pack.  

In order to reduce the statistical noise in the data we average over the six experiments. This is done by binning the x-axis (oscillation period) into equal sized regions (which are indicated on the inset) and averaging over all the measurements inside each bin. The averaged volume faction (for each bin) is plotted in the main graph in Figure~\ref{VF}. Again the heavy red line indicates the average volume fraction of the initial sample (i.e. before shaking), while the red dashed lines indicate experimental error in this measurement.

In each case, the biggest change in the volume fraction occurs after the first 300 seconds of shaking at the maximum frequency. Shaking the sample again, using a reduced frequency, can compact the sample further but the change at each such subsequent step is not as dramatic as initial change. Nevertheless, the system continues to compact linearly until at low frequencies oscillations the compaction saturates.

Figure~\ref{snaps}b shows the state of the grain column after the first 300s of shaking at the highest  frequency. It can be seen that the tracer grains (coloured white, blue and green) are almost completely redistributed, randomly, through the column and indeed at this high frequency we observe a convection like motion of the grains during shaking. Remarkably, when the shaking is stopped we find that the grains near the surface of the tube are, on average, aligned with their major axes along the axis of the tube. Further compaction does not produce an appreciable change in this surface texture, Figure~\ref{snaps}c shows the state of the system once the entire shaking protocol is finished. It is not clear if this nematic like ordering of the grains persists into the interior of the column or if it is an effect restricted to the surface. A similar orientational ordering has been observed in the compaction of rods, see \cite{villarruel, vandewalle}.

In contrast to the Firth grain, our experiments with oilseed grains only achieved a marginal compaction from an initial sample with a volume fraction of 0.50 to 0.53. This compaction was occurred entirely  after shaking the initial loose packed sample at maximum frequency. Subsequent shaking at lower frequencies did not yield further compaction. A more complete understanding of this difference between the rounded oil seed grains and the elongated Firth grains would have to take into account not only the morphology of the grains but also friction between grains, their shape and the influence on packing density due to the shape and size of the container.  

\subsection{Conclusions}

We have shown that on average a 6.6\% increase in packing density of a column of Firth grain can be achieved by shaking the sample according to a simple protocol (in one of the experiments this is increase is as large as 10\%). This is a strong indication that the current method used for measuring the HLW is inadequate and may largely give  random results. Clearly, a more robust method for characterising bulk grain density is needed and as such this is a topic that merits further study. 

Such future work would take into account the role of friction and cohesion between grains (which along with grain shape have been shown to play a major role in granular compaction dynamics \cite{vandewalle}), the effect on packing density due to the shape of the container and also an accurate account of the peak acceleration experienced by the system during shaking is needed. An exploration of different shaking protocols would also prove insightful.

Finally, we hope that studies of this kind may suggest methods to increase the density of stored grain. Even small improvements may, due to the economies of scale involved, lead to significant efficiencies in grain transport.

\section{Acknowledgements}
 
Participation by I.G. was part of the Quoats project funded through the Defra Sustainable Arable LINK Programme. Thanks are also due to Dr C Howarth for raising our interest in oat seed characteristics. J.W. acknowledges financial support through the joint IBERS-IMAPS summer studentship programme. We thank Mr. S. Fern, Mr. M. Gunn and Dr. D. Langstaff for their invaluable support with the experimental set-up.


\label{lastpage}

\end{document}